\documentclass[
 reprint,
 showpacs, preprintnumbers,
 amsmath, amssymb,
 aps,
 prl,
 floatfix,
 superscriptaddress
]{revtex4-2}

\usepackage{amsmath}
\usepackage{graphicx}
\usepackage{dcolumn}
\usepackage{bm}
\usepackage{bbm}

\usepackage[utf8]{inputenc}
\usepackage[T1]{fontenc}
\usepackage{mathptmx}
\usepackage{textgreek}
\usepackage{upgreek}
\usepackage{float}
\usepackage{hyperref}

\usepackage{color}

\usepackage{natbib}

\begin{document}

	\preprint{}

\title[JJ Diode]{Effect of Rashba and Dresselhaus spin-orbit coupling on  supercurrent rectification and magnetochiral anisotropy of ballistic Josephson junctions} 


\affiliation{Institut f\"ur Experimentelle und Angewandte Physik, University of Regensburg, 93040 Regensburg, Germany}
\author{C.~Baumgartner}
\author{L.~Fuchs}
\affiliation{Institut f\"ur Experimentelle und Angewandte Physik, University of Regensburg, 93040 Regensburg, Germany}
\author{A.~Costa}
\author{Jordi Pic\'{o}-Cort\'{e}s}
\affiliation{Institut f\"ur Theoretische Physik, University of Regensburg, 93040 Regensburg, Germany}
\author{S.~Reinhardt}
\affiliation{Institut f\"ur Experimentelle und Angewandte Physik, University of Regensburg, 93040 Regensburg, Germany}

\author{S.~Gronin}
\author{G.~C.~Gardner}
\affiliation{Microsoft Quantum Purdue, Purdue University, West Lafayette, Indiana 47907 USA}
\affiliation{Birck Nanotechnology Center, Purdue University, West Lafayette, Indiana 47907 USA}
\author{T.~Lindemann}
\affiliation{Department of Physics and Astronomy, Purdue University, West Lafayette, Indiana 47907 USA}
\affiliation{Birck Nanotechnology Center, Purdue University, West Lafayette, Indiana 47907 USA}

\author{M.~J.~Manfra}
\affiliation{Birck Nanotechnology Center, Purdue University, West Lafayette, Indiana 47907 USA}
\affiliation{Microsoft Quantum Purdue, Purdue University, West Lafayette, Indiana 47907 USA}
\affiliation{Department of Physics and Astronomy, Purdue University, West Lafayette, Indiana 47907 USA}
\affiliation{School of Materials Engineering, Purdue University, West Lafayette, Indiana 47907 USA}
\affiliation{School of Electrical and Computer Engineering, Purdue University, West Lafayette, Indiana 47907 USA}

\author{P.~E.~Faria~Junior}
\author{D.~Kochan}
\author{J.~Fabian}
\affiliation{Institut f\"ur Theoretische Physik, University of Regensburg, 93040 Regensburg, Germany}

\author{N.~Paradiso}\email{nicola.paradiso@physik.uni-regensburg.de}
\author{C.~Strunk}
\affiliation{Institut f\"ur Experimentelle und Angewandte Physik, University of Regensburg, 93040 Regensburg, Germany}
%


\begin{abstract}	
Simultaneous breaking of inversion- and time-reversal symmetry in Josephson junction leads to a possible violation of the $I(\varphi)=-I(-\varphi)$ equality for the current-phase relation. This is known as anomalous Josephson effect and it produces a phase shift $\varphi_0$ in sinusoidal current-phase relations. In ballistic Josephson junctions with non-sinusoidal current phase relation the observed phenomenology is much richer, including the supercurrent diode effect and the magnetochiral anisotropy of Josephson inductance. In this work, we present measurements of both effects on arrays of Josephson junctions defined on epitaxial Al/InAs heterostructures.  We show that the orientation of the current with respect to the lattice affects the magnetochiral anisotropy, possibly as the result of a finite Dresselhaus component. In addition, we show that the two-fold symmetry of the Josephson inductance reflects in the activation energy for phase slips.

\end{abstract}


\maketitle

Charge transport in superconductors is driven by the phase gradient of the condensate wavefunction. In the exemplary case of Josephson junctions~(JJs), this leads to a well-defined current-phase relation (CPR), which describes how the current $I$ depends on the phase difference $\varphi$ between the leads~\cite{RMPGolubov}. The CPR critically depends on the symmetries of the system: in particular \textit{either} time-reversal \textit{or} parity symmetry require that $I(\varphi)=-I(-\varphi)$. As a consequence, $I(0)=0$ and the CPR can be written as a Fourier series of sine terms only, $I=\sum_nb_n\sin (n\varphi)$.  If the temperature is close to the critical temperature $T_c$, or if the junction is relatively opaque (tunnel limit), then  the CPR reduces to a sinusoidal relation, $I=I_0\sin \varphi$. 

To obtain a finite current at  zero phase (and vice versa) it is necessary to break the equivalence between the leads (given by the space inversion symmetry) and simultaneously the time-reversal symmetry. This is called the anomalous Josephson effect~\cite{Bezuglyi2002,Buzdin2007,Buzdin2008,Reynoso2008,Reynoso2012,Yokoyama2014,Shen2014,Konschelle2015}.   The effect  is possible in noncentrosymmetric superconductors with large spin-orbit interaction (SOI) in the presence of a magnetic field. In such systems theory predicts  the possibility to have $I(\varphi)\neq -I(-\varphi)$. In the case of a sinusoidal CPR, this is equivalent to a finite phase offset $\varphi_0$, so that $I=I_0\sin (\varphi +\varphi_0)$. Such $\varphi_0$-junctions have been experimentally demonstrated in several systems~\cite{Szombati2016,Assouline2019,Mayer2020b,Strambini2020}.  

Noncentrosymmetric superconductors in a magnetic field also show magnetochiral anisotropy (MCA) effects, which arise when certain physical quantities display correction terms linear both in current and magnetic field~\cite{RikkenPRL2001,RikkenPRL2005,RikkenPRB2019,legg2021giant,MengAPL2021,NagaosaPRL2016,wang2021gigantic}. The first quantity to display MCA is the resistance in the fluctuation regime of a superconductor near $T_c$~\cite{Hoshino2018,Tokura2018,Wakatsuki2017,Yasuda2019,Itahashi2020,Ideue2020,Zhang2020,Itahashi2020}. In a recent work~\cite{baumgartner2021josephson}, we have found that the kinetic inductance of a superconductor as well shows the MCA effect. For JJs this means that the Josephson inductance $L$ can be written as
\begin{equation}
L=L_0[1+\gamma_L \hat{e}_z(\vec{B}\times\vec{I})], 
\label{eq:gammaell}
\end{equation}
where $\gamma_L$ is the MCA coefficient for the inductance.  As a consequence of the MCA, the critical currents for the two directions of the supercurrent flow differ, leading to a finite current interval where the resistance is zero only for one bias polarity. This is a  supercurrent diode effect (first envisioned in Ref.~\cite{HuPRL2007}), which has been first  demonstrated in bulk superconductors~\cite{Ando2020} and then in JJs~\cite{baumgartner2021josephson}. In both cases, the SOI was of Rashba-type and the supercurrent rectification driven by the in-plane field perpendicular to the current. Very recent reports~\cite{bauriedl2021,shin2021} have demonstrated the supercurrent diode effect in systems with valley-Zeeman SOI (where the rectification is driven by the out-of-plane field, as expected from theory~\cite{he2021arx}), in type II Dirac semimetal~\cite{palparkin2021}, or in magic-angle twisted bilayer~\cite{diezmerida2021magnetic} or trilayer~\cite{lin2021zerofield,scammell2021theory} graphene. In turn, such intriguing experimental evidences have stimulated a number of theoretical studies on non-reciprocal supercurrent in exotic systems~\cite{scammell2021theory,zinkl2021symmetry,Fu2022,zhang2021general,halterman2021supercurrent,watanabe2021nonreciprocal,He2022}. 
As discussed in our previous work~\cite{baumgartner2021josephson}, in JJs with Rashba SOI, the supercurrent diode effect is strictly related to the $\varphi_0$ shift discussed above. The MCA and supercurrent diode generalize the anomalous Josephson effect to the case of nonsinusoidal (i.e., skewed) CPRs.

In this work, we study  the supercurrent diode effect and the MCA in JJ arrays  with large Rashba SOI. We present experiments that complement the main observations reported in Ref.~\cite{baumgartner2021josephson}: we show results for different lattice orientations, which we use to estimate the Dresselhaus contribution to the SOI; we study the MCA for the activation energy of thermally activated phase slips, which we connect to the anisotropy of the inductance. Our results provide a useful illustration of the role of SOI on the CPR. 

\section*{Ballistic SNS junctions in epitaxial AL/InAs heterostructures}

A clean superconductor with synthetic Rashba interaction can be produced combining an epitaxial Al film and a high-mobility 2D electron gas (2DEG) confined in a InGaAs/InAs quantum well. If the barrier between 2DEG and Al film is transparent, then the 2DEG former will be proximitized by the superconducting Al,   leading to a Rashba 2D superconductor. The combination of superconductivity and SOI is at the basis of topological superconductivity.  This material platform has therefore been mainly developed  by the community studying Majorana modes and topological superconductivity. 

Our devices are fabricated starting from a heterostructure whose top layers are Al (7~nm)/In$_{0.2}$Ga$_{0.8}$As (10~nm)/InAs (2DEG) (7~nm)/In$_{0.2}$Ga$_{0.8}$As (4~nm) (see the Supplementary Material of Ref.~\cite{baumgartner2020} for the full layer sequence).  
To obtain JJs of finite width $W$ we define a mesa using a phosphoric acid-based wet etching procedure. The chosen width $W$ is the result of a compromise between a sufficiently high Josephson coupling (increasing with $W$) and a measurable Josephson inductance (decreasing with $W$). In our experiments this leads to widths of the order of few micrometers, which correspond to a critical current of several microamperes.

\begin{figure*}[tb]
\includegraphics[width=2\columnwidth]{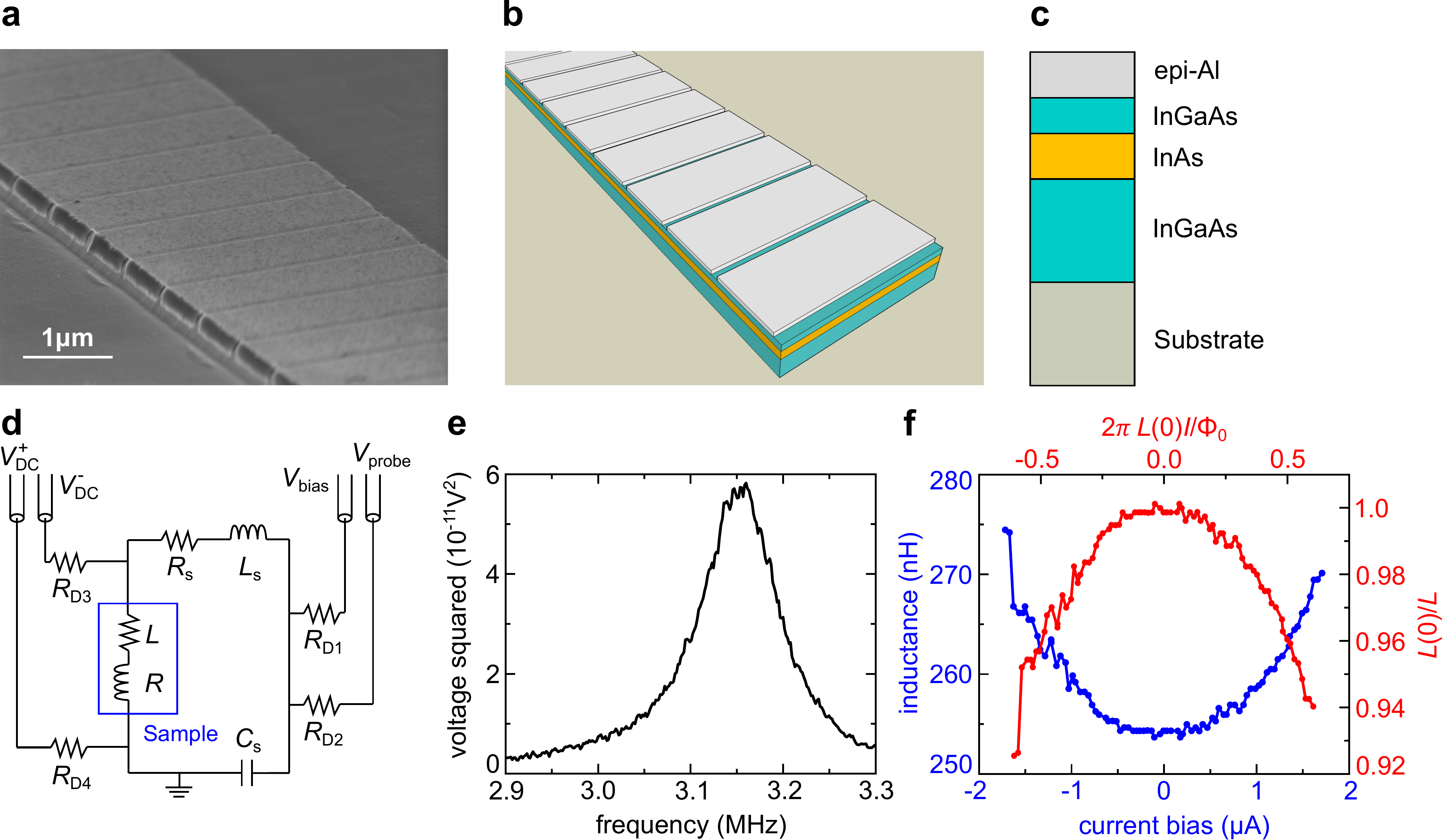}
\caption{(a) Scanning electron micrograph of a portion of the Josephson junction arrays under study (sample 3). The 3.3 \textmu m-wide mesa is fabricated by deep etching, while the gaps between the top Al islands are obtained by selective etching.   (b) Corresponding scheme of the structure. The yellow part highlights the quantum well where the 2D electron gas is located. (c) Top layer sequence for the heterostructure. As substrate, we indicate the remaining layers, not relevant for the transport. The complete sequence can be found in Ref.~\cite{baumgartner2020}. (d) Scheme of the cold RLC circuit located next to the sample. (e) Resonance curve measured for sample 3 at $T=$100~mK, and at zero applied DC bias and field. (f) Measured $L(I)$ curve for the same sample. Each $L$ value is deduced from the center frequency of the corresponding RLC resonance spectrum.}
\label{fig:firstfig}
\end{figure*}

SNS junctions are obtained by selectively etching Al to form 100~nm-long gaps separating the remaining rectangular Al islands.  A SEM picture of the device is shown in Fig.~\ref{fig:firstfig}(a), while a sketch of it is depicted in Figs.~\ref{fig:firstfig}(b,c).
 The selective etching is by far the most critical step of the fabrication process. The disorder introduced into the shallow 2DEG of the exposed InAs regions must be minimized in order to obtain ballistic junctions with high transparency.

\section*{Cold RLC resonators for Josephson inductance measurements in the MHz regime}

DC transport measurements provide only partial information about single JJs. 
For instance, the Josephson coupling between the leads is deduced from the critical current --- an interesting situation where an \textit{equilibrium} quantity is deduced from AC transport measurements. The CPR  is not accessible without making use of a SQUID geometry in perpendicular magnetic field. Josephson coupling and CPR can instead be directly accessed in single junctions by measuring its Josephson inductance, clearly with AC measurements. For example, given the CPR relation $I=I_0 f(\varphi)$ (where $I_0$ is the relevant current scale and $f$ is a  $2\pi$ periodic function) the Josephson inductance immediately emerges from the ratio between Josephson voltage and time derivative of the CPR
\begin{equation}
    L(\varphi) = \frac{V}{\frac{d}{dt}I}=\frac{\Phi_0\frac{d \varphi}{dt}\frac{1}{2\pi}}{\frac{d}{dt}[I_0f(\varphi)]}= \frac{\Phi_0}{2\pi I_0 f^{\prime} (\varphi)}.
    \label{eq:JosInd}
\end{equation}
In a simple junction without loop, it is the current and not the phase that is controlled, therefore it is convenient to integrate $\dot{\varphi}=2\pi L\dot{I}/\Phi_0$ to obtain the inverse CPR $\varphi(I)$:
\begin{equation}
    \varphi(I)-\varphi(0)=\frac{2\pi}{\Phi_0}\int_0^I L(I^{\prime})dI^{\prime},
    \label{eq:intphase}
\end{equation}
where $L(I)$ is the measured quantity. Therefore, the Josephson inductance as a function of the current is proportional to the derivative of the inverse CPR $\varphi(I)$. 

The difficulties in the measurement of the Josephson inductance are related to the fact that it is typically much smaller than the inductance of the cryostat cables. To decouple the sample from the external cabling, it is possible to embed the sample in a low resistance resonator decoupled from the external leads by resistors. The resonance frequency will then directly provide the inductance if the $Q$ factor is above unity. The circuit scheme of the RLC resonator used in this work is shown in Fig.~\ref{fig:firstfig}(d). The $Q$ factor of the loop is approximately given by the formula for series RLC tank, $Q=\frac{1}{R_l}\sqrt{\frac{L_l}{C_s}}$. Here $R_l$ and $L_l$ are, respectively, the total resistance (i.e., sample resistance $R$ plus external circuit resistance in series $R_s$) and total  inductance of the loop (i.e., sample inductance $L$ plus external circuit inductance in series $L_s$), while $C_s$ is the series capacitance, given by an external capacitor. 
The choice of the working point for the frequency is crucial. For a series RLC, the higher the frequency, the higher the sensitivity. On the other hand, at very high frequency measurements in the presence of magnetic fields are difficult since emerging dissipation would immediately damp the resonance. Moreover, at high frequency (rf regime close to the plasma frequency) the physics of a JJ is not equivalent to that in DC.  At very high frequency, one must take into account transmission line  resonance on the very sample. 

For our measurements we have chosen to operate in the MHz range. This frequency regime allows us to operate under large magnetic fields without significant damping of the resonance. Also, this frequency range is below any relevant physical threshold for the JJs under study (plasma frequency $\omega_P\approx 240$~GHz, first transmission line mode for the array $\omega_0=250$~MHz), so that for any practical purpose we are operating in quasi-DC regime. Figure~\ref{fig:firstfig}(e) shows a typical resonance spectrum measured by lock-in, whose center frequency directly provides the inductance. The Josephson inductance is obtained by subtracting the external inductance of the circuit, which has been determined in a dedicated calibration session. Our typical inductance measurement consists in measuring RLC spectra as a function of control parameters, e.g., the DC current as in the measurement of $L(I)$ shown in Fig.~\ref{fig:firstfig}(f). The inductance is deduced from the resonance frequency, since the external resistance $R_s$, capacitance $C_s$ and inductance $L_s$ in series to the sample [see circuit scheme in Fig.~\ref{fig:firstfig}(d)] are known.

This method (which is an adaptation, with modern electronics, of the experiment in Ref.~\cite{Meservey1969}) makes it also possible to accurately extract the sample resistance $R$  via the resonance quality factor $Q$. However, for our circuit parameter, the $Q$ factor is suppressed already for resistances of the order of 1~\textOmega , which is roughly four orders of magnitude less than the normal resistance of our samples. This means that the inductance measurements shown here are all conducted deep in the superconducting regime, where the resistance is a very small fraction of the normal state resistance.

As shown below, the typical inductance of a 3~\textmu m-wide and 100~nm-long JJ is of the order of 100~pH. In the MHz regime we operate in, this inductance is below the  resolution limit of our electronics. Therefore, instead of a single junction, we measure an array of thousands junctions. If their spacing is sufficient to exclude mutual (e.g., magnetic) interaction, they will behave as a set of inductors in series, i.e., the measured inductance will reflect the sum of the ensemble. This configuration has advantages and disadvantages. The disadvantage  is that the critical current (or field) is  set by the weakest junction. When that value is reached the emerging resistance is enough to damp the resonance. For this reason, measured $L(I)$ curves, as e.g.~that  shown in Fig.~\ref{fig:firstfig}(f), terminate before the expected divergence at the critical current. Working with arrays has also crucial advantages beyond the obvious increase in sensitivity.  In fact, in large JJ arrays imperfections in single JJs are unimportant, since only the average behavior is measured. If a weaker junction is present, its inductance will be higher than the typical one, but it will hardly affect the total inductance given by thousands of JJs. This is true as long as the current is below the reduced  critical current value for the weakest junction, as discussed above.

\section*{Characterization of ballistic Josephson junctions in Rashba 2DEGs}

The CPR of short ballistic SNS junctions at finite temperature $T$ can be described by the complete Furusaki-Beenakker formula~\cite{Furusaki_1991,BeenakkerPRL91}
\begin{equation}
I(\varphi)=I_0f(\varphi)=I_0\frac{\bar{\tau}\sin \varphi \tanh \left[\frac{\Delta^{\ast}}{2k_BT}\sqrt{1-\bar{\tau}\sin^2\left(\frac{\varphi}{2}\right)} \right]}{2\sqrt{1-\bar{\tau}\sin^2\left(\frac{\varphi}{2}\right)}},
\label{eq:cprshortbalFB}
\end{equation}
where $\bar{\tau}$ it the transmission coefficient and $\Delta^{\ast}$ is the effective gap at the leads. In our case $\bar{\tau}$ refers to the \textit{average} transmission coefficient, while $\Delta^{\ast}$ referes to the induced gap in the 2DEG region just underneath the epitaxial Al film. The characteristic current $I_0$ (which coincides with the critical current only for $\bar{\tau}\rightarrow 1$ and $T\rightarrow 0$) is given by
\begin{equation}
I_0 = \frac{e\Delta^{\ast}}{\hbar}\,N,
\label{eq:I0}
\end{equation}
where $N$ is the number of spin-degenerate transverse modes in the channel. To characterize our junctions we need three parameters, namely $I_0$, $\bar{\tau}$ and $\Delta^{\ast}$. The first two can be obtained from a $L(I)$ measurement at $T/T_c\ll 1$. In particular, $\bar{\tau}$ is determined from the curvature of the graph of $L(0)/L$ versus $2\pi L(0)I/\Phi_0$, which in the low temperature limit depends only on $\bar{\tau}$~\cite{baumgartner2020}, see red curve in Fig.~\ref{fig:firstfig}(f).

The transmission coefficient strictly depends on the quality of the selective Al etching defining the weak link. In our best JJ arrays, we obtained average transmission close to unity, e.g., $\bar{\tau} = 0.94$ in Ref.~\cite{baumgartner2020}. If $\bar{\tau}$ is found (and thus the low temperature limit of the CPR), $I_0$ can then be calculated from $L(0)=\Phi_0/[2\pi I_0 f^{\prime}(0)]$. 

The characterization of $\Delta^{\ast}(T)$ requires, instead, data at finite temperature, as it is evident from Eq.~\ref{eq:cprshortbalFB}. It is important to notice that for an epitaxial Al/InAs 2DEG bilayer, the temperature dependence of the induced gap  $\Delta^{\ast}(T)$ differs from that predicted by BCS ~\cite{Chrestin1997,Aminov1996,Schaepersbook,KjaergaardPRAPPL17}. More precisely $\Delta^{\ast}(T)$ depends on both the BCS-like gap of the Al film and on the coupling coefficient $\gamma_B$ between Al film and 2DEG (see, e.g., Eq.~17 in Ref.~\cite{Aminov1996} or Eq.~S7 in Ref.~\cite{baumgartner2020}) which can be determined by fit. However, in the low temperature limit this dependence on $\gamma_B$ is weak, therefore the extracted $\Delta^{\ast}(0)=130$~\textmu eV value is independent of theory. In fact, when plugged into Eq.~\ref{eq:I0} it provides a number of channels very close to that extracted from the Sharvin resistance~\cite{baumgartner2020}.

The robust determination of the number of transverse channels in a 3.15~\textmu m-wide conductor allows us to deduce the Fermi wavelength $\lambda_F=33$~nm. To extract the Fermi velocity, an estimate for the effective mass is needed. For bulk InAs the best estimate~\cite{Vurgaftman2001} is $m^{\ast}=0.026m_0$, where $m_0$ is the electron mass.In quantum wells, owing to confinement, the effective mass is renormalized~\cite{Yang1993,Prior2005,Yuan2020}. In quantum wells with similar layer composition, but with narrower quantum well (4~nm as opposed to 7~nm in our structures), the effective mass was measured to be $m^{\ast}=0.04m_0$~\cite{Yuan2020}. Since in our wafers the confinement is less pronounced, we expect the effective mass to be in between $0.026 m_0$ (bulk InAs) and 
$0.04 m_0$ (Ref.~\cite{Yuan2020}), i.e., a value near $m^{\ast}=0.03 m_0$. With this assumption, we deduce a Fermi velocity $v_F=7.31\cdot  10^5$~m/s. The Fermi velocity, together with Al superconducting gap ($\Delta_{\text{Al}}=220$~\textmu eV~\cite{baumgartner2020}), allows us to estimate an important parameter for SNS junctions, namely, $\lambda=\mathcal{L}\Delta_{\text{Al}}/(\hbar v_F)=0.046$, where $\mathcal{L}=100$~nm is the junction length~\cite{Metzger2021}. The fact that  $\lambda$ is so much smaller than unity implies that our junctions can be considered to be deeply in the short junction limit. 
It allows us to discriminate the above mentioned renormalization of the effective gap~\cite{KjaergaardPRAPPL17} from the simple reduction of the ABS-energy because of a non-vanishing length of the junction.

\section*{SNS junctions with strong spin-orbit coupling: $\varphi_0$ and supercurrent diode effect}

\begin{figure*}[tb]
\includegraphics[width=\textwidth]{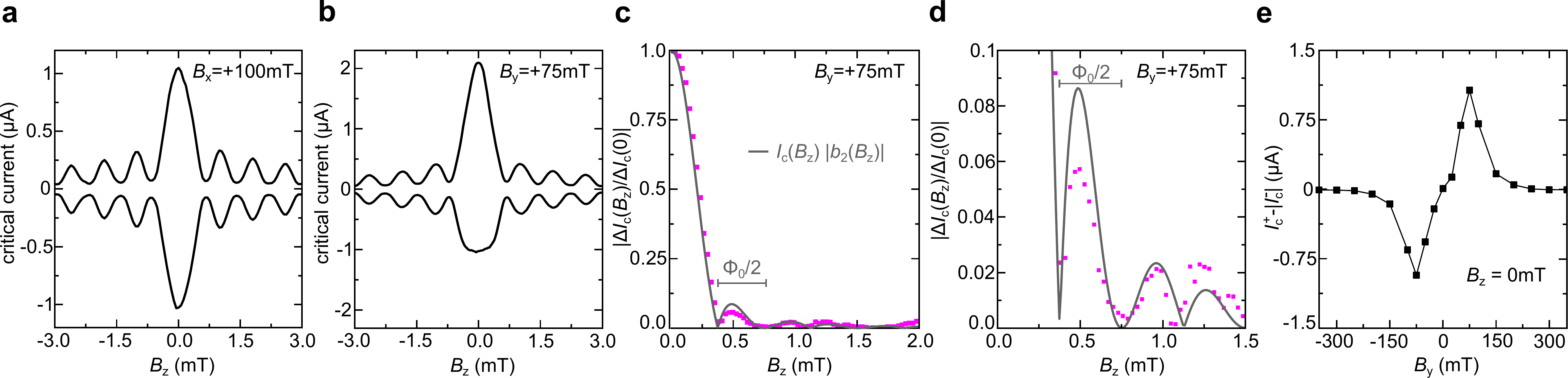}
\caption{(a) Fraunhofer pattern for sample 1, measured applying an in plane field $B_{ip}=100$~mT \textit{parallel} to the current direction. (b) The same measurement for $B_{ip}=75$~mT applied \textit{perpendicular} to the current direction. (c) Measured (at $B_{ip}=75$~mT) absolute value of the difference $|\Delta I_c|$ between the critical currents for positive and negative bias  (symbols), and calculated product $I_c\cdot b_2$ of critical current and second Fourier coefficient (gray line). Both curves are plotted as a function of the out-of-plane field $B_z$
and normalized to the $B_z=0$ value. (d) Zoom-in that highlights the oscillations of the critical current difference with flux period $\Phi_0/2$. (e) $\Delta I_c$ versus $B_{ip}$ at $B_z=0$.
} 
\label{fig:FP}
\end{figure*}

 In this section we will discuss the consequences of the ballistic character on the physics of JJs in the presence of large SOI.
Several recent works~\cite{Szombati2016,Assouline2019,Mayer2020b,Strambini2020} have demonstrated that, in the presence of Rashba SOI and broken time-reversal symmetry, a JJ might show an anomalous Josephson effect, i.e., a $\varphi_0$ shift of the CPR. In multimode conductors, the $\varphi_0$ shift is mainly determined by the channels with low Fermi velocity. 

In a $\varphi_0$-junction, a purely sinusoidal CPR is still antisymmetric around the ($\varphi_0$-shifted) zero current point, i.e., if $\phi\equiv \varphi-\varphi_0$, then $I(\phi)=-I(-\phi)$. Thus, the positive and negative branch of the CPR are equal and opposite, and the CPR inflection point still occurs at $I=0$. In our experiments, we show that if the junction is ballistic and an in-plane field is applied perpendicular to the supercurrent direction, then  SOI will not simply lead to a $\varphi_0$ shift, but will effectively distort the CPR in an asymmetric way, so that positive and negative branches will differ. As a consequence, positive and negative critical currents will differ, giving rise to the   supercurrent diode effect. 

In the experiments reported in Ref.~\cite{baumgartner2021josephson}, we show that the main effect of an in-plane field perpendicular to the current  consists in the addition of a cosine term to the CPR, which only contains sine terms in the absence of the magnetic field. 
If the unperturbed CPR is purely sinusoidal (low $ \overline{\tau} $ limit), the addition of a cosine produces a shifted sine, i.e., a phase shift $\varphi_0$. However, if higher harmonics in the CPR are not negligible, the situation is more complex. Let us consider
 first the Fourier expansion of a generic skewed CPR
\begin{equation}
    I(\varphi)=\sum_n b_n \sin (n\varphi).
    \label{eq:fourier}
\end{equation}
In the presence of magnetic field (and thus anomalous Josephson effect) each $n$-term would acquire its own $\varphi_{0,n}$ shift. The determination of each $\varphi_{0,n}$ is nontrivial~\cite{Yokoyama2014}, and in general $\varphi_{0,n} \neq  n\varphi_0$, where $\varphi_{0}$ is some common phase shift. Therefore, the resulting CPRs will not be merely shifted. Instead, each $\varphi_{0,n}$ shift will be equivalent to the addition of a $a_n \cos (n\varphi)$ term in the Fourier series. In the Furusaki-Beenakker CPR, the Fourier coefficients $b_n$ are exponentially suppressed with $n$ (see Supplementary Information in Ref.~\cite{baumgartner2021josephson}), therefore, in a rough  approximation, only the first terms will be important. Keeping only the leading terms in the approximation 
\begin{equation}
    I(\varphi) \approx  b_1 \sin (\varphi)  +a_1 \cos (\varphi) + b_2 \sin (2\varphi),
    \label{eq:fourierapprox}
\end{equation}
where  $a_1$ is proportional to both magnetic field and Rashba SOI strength, while $b_2$ mainly determine the skewedness. The MCA effect requires both $a_1$ and $b_2$ to be nonzero, while the simple anomalous shift $\varphi_0$ only requires the former.

The supercurrent diode effect can be measured with standard DC transport experiments. Figure~\ref{fig:FP}(a) shows the Fraunhofer pattern measured in sample 1 in the presence of an in-plane field ($B_x=100$~mT) directed parallel to the supercurrent direction. For this field alignment there is no MCA effect: positive and negative critical currents are the same, i.e., the graph is symmetric around the abscissa axis. In contrast, in the presence of an in-plane field $B_{ip}=75$~mT perpendicular to the current, the critical currents for opposite polarities are different, see Fig.~\ref{fig:FP}(b). We stress that critical current values in both Fig.~\ref{fig:FP}(a) and Fig.~\ref{fig:FP}(b) were obtained by sweeping the current from zero to finite (either positive or negative) bias. Interestingly,  when the supercurrent diode effect is enabled [Fig.~\ref{fig:FP}(b)], the critical current difference is pronounced only for small values of $B_z$, as visible in Fig.~\ref{fig:FP}(c): this figure shows (symbols) the absolute value of the critical current difference as a function of $B_z$, normalized to the $B_z=0$ value. We notice also that such difference oscillates with $B_z$ with a flux period of $\Phi_0/2$, as it can be seen in the zoomed graph in Fig.~\ref{fig:FP}(d). The peculiar $B_z$ dependence can be captured by the product of the critical current and the first higher harmonic Fourier coefficient $b_2$ of the Fourier expansion of the Furusaki-Beenakker CPR. The former term contains the envelope of all the harmonics producing the Fraunhofer pattern, while the latter contains the most relevant term for the skewedness of the CPR, which as explained above, determines the diode effect. The curve, calculated from Eq.~\ref{eq:cprshortbalFB} with parameters extracted from the experiment~\cite{baumgartner2020,baumgartner2021josephson} [solid line in Fig.~\ref{fig:FP}(d)]   nicely matches the experimental data, including the alternating sequence of cusp-like and quadratic minima. This clearly demonstrates that in short-ballistic JJs the diode effect is mainly determined by the first higher harmonic $b_2$ above the fundamental term.

Finally, for $B_z=0$ we can extract the $B_{ip}$ dependence of the diode effect, depicted in Fig.~\ref{fig:FP}(e). Up to about $B_{ip}\approx 80$~mT, the dependence is nearly linear, as expected by a magnetochiral effect, see Eq.~\ref{eq:gammaell}. Above this threshold, the critical  current asymmetry rapidly decreases, indicating that pair-breaking is at work. Interestingly, the diode effect is more fragile than bare superconductivity, since it relies on higher harmonics of the CPR, which are suppressed well before the fundamental term. Thus, at sufficiently high field, one still observes finite critical current but no supercurrent diode effect.

\section*{Impact of lattice orientation on magnetochiral anisotropy}

\begin{figure*}[tb]
\includegraphics[width=\textwidth]{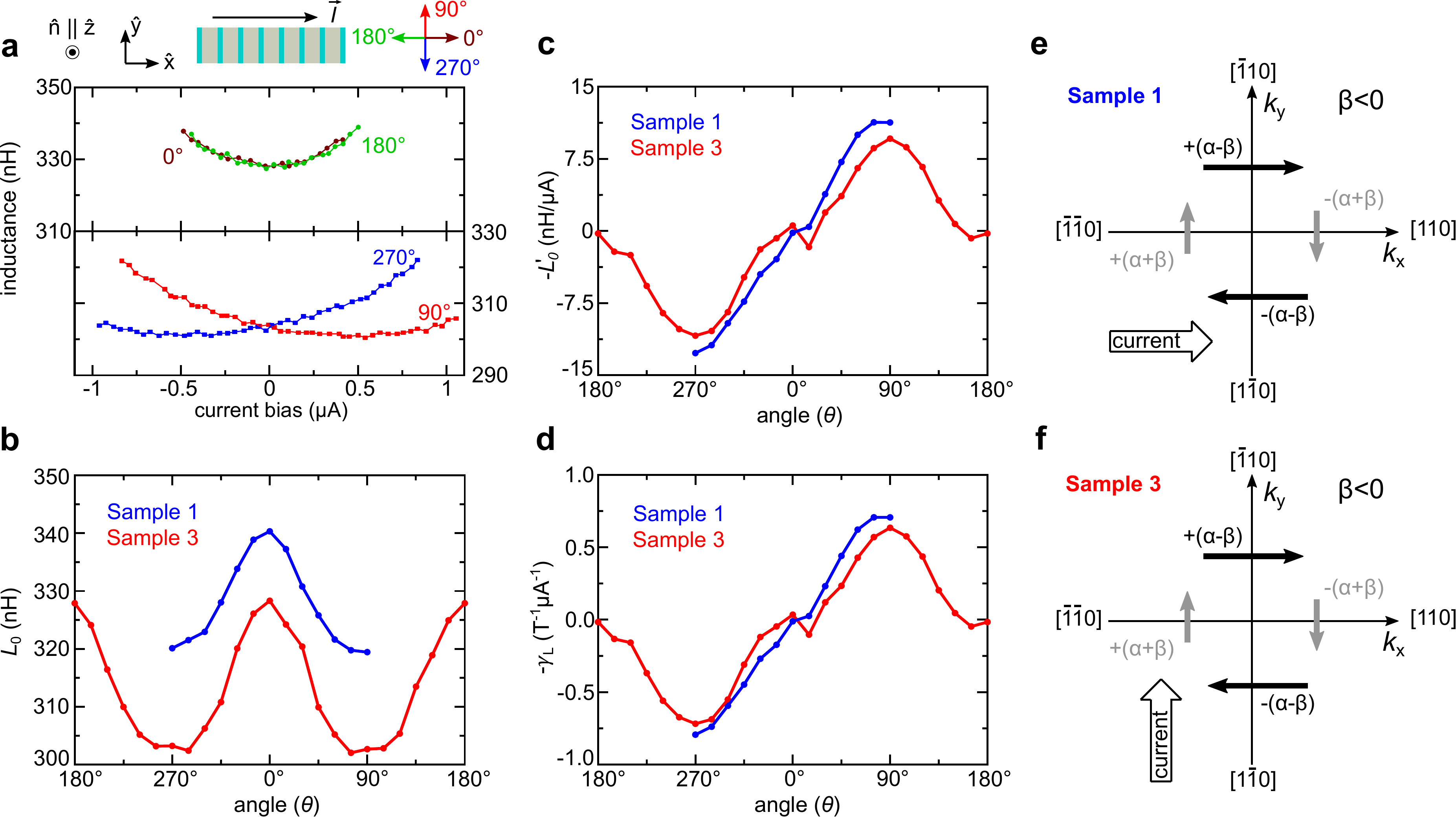}
\caption{(a) The top sketch shows the mutual orientation of the vectors current $\vec{I}$, in-plane field $\vec{B}_{ip}$, and $\hat{n}$ (unit vector perpendicular to the surface and pointing to the top) for $\theta=0^{\circ}$ (brown), $\theta=90^{\circ}$ (red), $\theta=180^{\circ}$ (green), and $\theta=270^{\circ}$ (blue). The graphs show the $L(I)$ curve for each value of $\theta$, with the same color code. (b) Constant term $L_0$ (see text) for the measured $L(I)$ curve, plotted as a function of $\theta$ for sample 1 (blue) and sample 3 (red).  In sample 1 (sample 3) the current is directed along the $[110]$  ($[1\bar{1}0]$) direction. (c)  Plot of $L^{\prime}_0$ as a function of $\theta$ for sample 1 and sample 3. (d) Plot of $\gamma_L (\theta)$. (e) Scheme showing the magnitude of the total (Rashba plus Dresselhaus) SOI field in sample 1, sketched for different $\vec{k}$  with respect to the current direction (horizontal). Here we assume $\beta<0$, which is the case for InAs quantum wells. (f) The same for sample 3. In sample 1 the current direction is directed along the $\vec{k}$ direction where the SOI field is the largest (Rashba and Dresselhaus add), while for sample 3 the current points to the $\vec{k}$ direction of least SOI field. 
 } 
\label{fig:Dresselhaus_Rashba}
\end{figure*}

As discussed above, the supercurrent diode effect is a direct consequence of the distortion of a skewed CPR produced by an in-plane field in the presence of a spin-split conduction band. The asymmetry between the positive and negative CPR branches implies that $(i)$  positive and negative critical currents are different, $(ii)$ the inductance is not an even function of $I$ anymore, owing to the  magnetochiral correction term linear in both current and field, see Eq.~\ref{eq:gammaell}. The inductance is proportional to the derivative of the inverse CPR, $\varphi(I)$, therefore  $(ii)$ is equivalent to a shift of the CPR inflection point from $I=0$ to a finite value $I=i^{\ast}$. The latter value can be experimentally determined from the minimum of the $L(I)$ curve. 

In our junctions the MCA is a small correction, therefore we can approximate the $L(I)$ curve as a parabola near zero current, i.e., $L(I)\approx L_0 + L^{\prime}_0 I + L^{\prime \prime}_0 I^2$. The CPR can then be characterized by the three coefficients $L_0$, $L^{\prime}_0$, and $L^{\prime \prime}_0$. In particular, it is $L^{\prime}_0$ that mostly determines the MCA. At finite in-plane field $B_{ip}$, one can extract the MCA coefficient $\gamma_L \equiv -2L^{\prime}_0/(L_0 B_{ip})$.  As shown in the $L(I)$ measurements plotted in Fig.~2 of Ref.~\cite{baumgartner2021josephson}, our experiments confirm that it is indeed the in-plane field component perpendicular to the current that determines the MCA. In fact, if the sample is rotated keeping constant the in-plane field magnitude and direction, then both $L^{\prime}_0$ and $\gamma_L$ display a sinusoidal dependence on the angle between $\vec{I}$ and $\vec{B}_{ip}$, with the maximum anisotropy occurring for $\theta=\pm 90^{\circ}$.

For a conductor with purely Rashba SOI, the direction of the current with respect to the underlying lattice is unimportant, since both spin-split Fermi surfaces are isotropic. For generic Rashba ($\alpha$) and Dresselhaus ($\beta$) SOI parameters, the spin-orbit field $\vec{\Omega}$  is defined such that the perturbative SOI term of the Hamiltonian is~\cite{Ganichev2004}  
\begin{equation}
    H_{SOI}= \vec{\Omega} \cdot \hat{\vec{\sigma}}=(\alpha-\beta)k_y \hat{\sigma}_x-(\alpha+\beta)k_x \hat{\sigma}_y . 
    \label{rq:RD}
\end{equation}
The magnitude of the MCA effect depends on the $\vec{\Omega}$ component parallel to the current, corresponding to a $k$-space direction  \textit{perpendicular} to the current [$k_y$ direction for our axis choice, see Fig.~\ref{fig:Dresselhaus_Rashba}(e,f)]. In the pure Rashba SOI case ($\beta =0$), the modulus of the pseudo-magnetic field $|\vec{\Omega}|$ is isotropic (its magnitude does not depend on the direction in the reciprocal space), and thus the particular mutual orientation of current and lattice is irrelevant.

The situation changes in the presence of a small Dresselhaus SOI component ($\beta \neq 0$). In this case, the total spin-orbit field $|\vec{\Omega}|$ is reduced (enhanced) for the $k$-direction where Rashba and Dresselhaus SOI fields are antiparallel (parallel).
As a result, a finite Dresselhaus component breaks the symmetry among different crystal directions. To verify the presence of a Dresselhaus component, we have fabricated an array (sample 3) which is, to the best of our ability, identical to the array used for the measurements reported above (sample 1). The only nominal difference is the orientation of the current with respect to the lattice axes: in sample 1 the current is directed along the [110] direction, while in sample 3 it is directed along the [$1\overline{1}0$] axis. We have then repeated the inductance MCA measurements, whose results are summarized in Fig.~\ref{fig:Dresselhaus_Rashba}(a). The $L(I)$ curves for different angles $\theta$ between $\vec{B}_{ip}$ and $\vec{I}$ in sample 3 are similar to those for sample 1 reported in Ref.~\cite{baumgartner2021josephson}. Using the same procedure described there, we can extract the $L_0$, $L^{\prime}_0$, and $L^{\prime \prime}_0$ coefficients; $L_0$ and $L^{\prime}_0$ are plotted as a function of $\theta$ in Fig.~\ref{fig:Dresselhaus_Rashba}(b) and (c), respectively. The blue (red) curve refers to sample 1 (sample 3).  First, we notice that the two $L_0$ coefficients are very similar (indicating a good reproducibility of the fabrication procedure) and both have a very weak angular dependence [notice the small range for the vertical axis in Fig.~\ref{fig:Dresselhaus_Rashba}(b)]. Second, the $L^{\prime}_0$ coefficients show small, but important, differences: the amplitude of the 90$^{\circ}$-270$^{\circ}$ excursion is larger in sample 1, while an anomalous plateau near $\theta=0^{\circ}$ is more pronounced in sample 3. From those values we can calculate $\gamma_L$ for the two samples, see Fig.~\ref{fig:Dresselhaus_Rashba}(d). From the amplitude of the quasi-sinusoidal curves, we deduce a ratio $r$ between the maximum $L_0^{\prime}$ for sample 3 (current parallel to [1$\bar{1}$0])  and that for sample 1 (current parallel to [110]). In our experiment we obtain $r=0.854$.

As discussed above, $r \neq 1$ can be attributed to a Dresselhaus SOI component. Numerical quantum transport simulations (computed with the KWANT package~\cite{Groth2014}, using the methodology and parameters as described in Ref.~\cite{baumgartner2021josephson}) found that in good approximation $r$ is a linear function of $ |\beta| $, more precisely $r \approx 1.004  -0.225 |\beta|$ with $\beta$ expressed in meV/nm units. 
On this basis, we can estimate a Dresselhaus parameter of $ \beta \approx - 0.67 $~meV/nm, which is approximately in line with the $ \vec{k} \cdot \vec{p} $~estimate reported in the Supplementary Information of Ref.~\cite{baumgartner2021josephson}.





\section*{Angle dependence of the thermal activation for phase slips}

In the previous sections we have discussed in detail the two-fold anisotropy induced by the in-plane field on $L_0^{\prime}$, and thus on the inflection point of the CPR, which is at the basis of MCA. We have also highlighted a weaker, but still evident anisotropy in $L_0$. A two-fold anisotropy in $L_0$ is expected to produce a similar anisotropy in the Josephson coupling $E_J=\hbar I_c/2e$ and, consequently, an anisotropic activation energy for phase slips in the junctions.  

The experiments discussed so far mainly focus on the deep superconducting regime at temperatures close to the base temperature of our dilution refrigerator, where $ T \ll \Delta^{\ast}/k_B$. For our JJ arrays the Josephson energy is much larger than the charging energy, thus in this regime we cannot detect any resistive phase slip effect. To investigate the angle dependence of the phase slip rate, we must work in a temperature regime closer to $T_c$. However, one must keep in mind that sample resistances larger than few ohms are not compatible with the resonator technique. The resonator can indeed be used to measure very small resistance changes via the $Q$ factor, but as long as the total resistance of the RLC tank is above few ohms (roughly 1~m\textOmega~per junction), the resonance is suppressed altogether. Hence, we studied phase slip rates by conventional DC  transport measurements.

\begin{figure*}[bt]
\includegraphics[width=\textwidth]{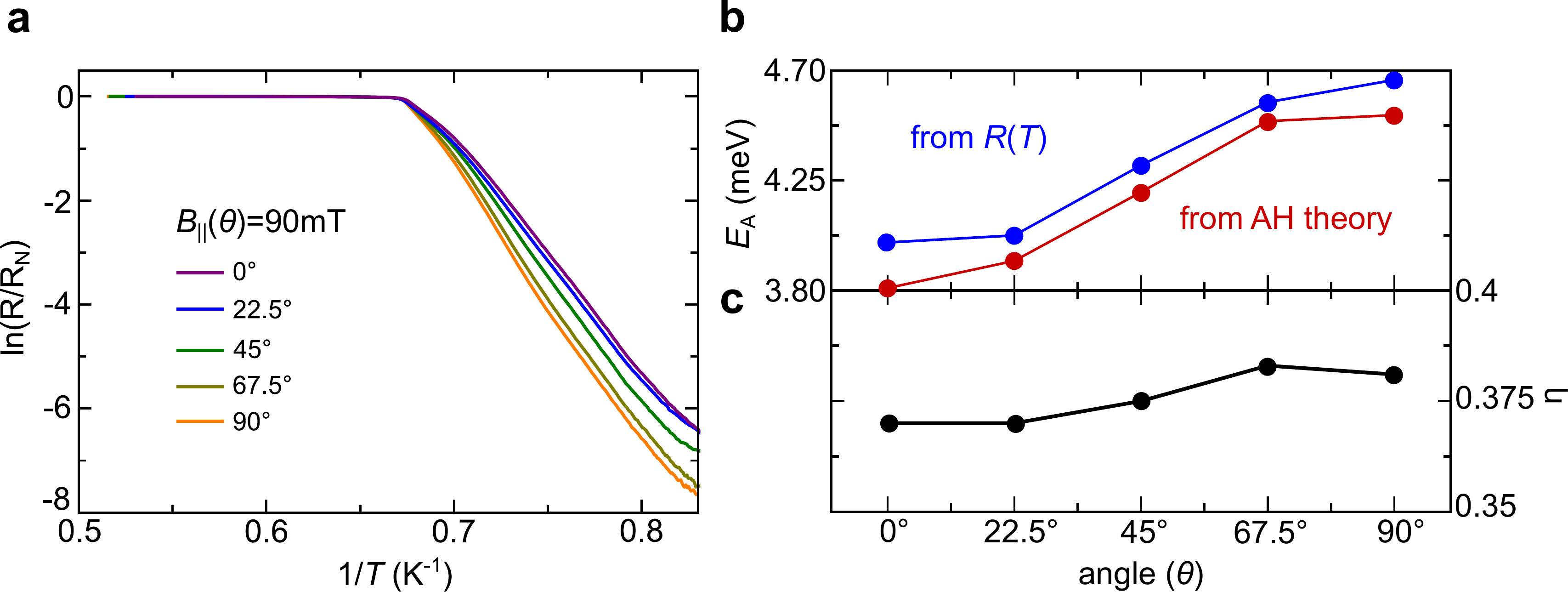}
\caption{(a) Arrhenius plot of the temperature dependence of the resistance $R(T)$, plotted for different angles $\theta$ between current and in-plane magnetic field. (b) Activation energy extracted from the linear part of the graph panel (a) (blue symbols), plotted together with twice the Josephson energy $2E_J$ calculated via the Ambegaokar-Halperin
theory (red, see text).
} 
\label{fig:activation_energy}
\end{figure*}

Figure~\ref{fig:activation_energy}(a) shows the  Arrhenius plot of the temperature-dependent resistance near $T_c$ in an in-plane field of 90~mT. Each curve refers to a different angle $\theta$ between in-plane field $\vec{B}_{ip}$ and current $\vec{I}$. The resistance is clearly  thermally activated with an activation energy that depends on the angle of the in-plane field. At the lowest temperature, there are deviations from the Arrhenius law, most probably due to 2-3  junctions with reduced $I_c$. From the linear part of the Arrhenius curve we can extract the activation energy, which is plotted versus $\theta$ in Fig.~\ref{fig:activation_energy}(b) (blue symbols). In the same curve (red symbols),  we show the corresponding  values of twice the Josephson energy, $2E_J$, calculated via the Ambegaokar-Halperin theory~\cite{AmbegaokarHalperin1969}, adapted to describe junctions with the non-sinusoidal CPR as in Eq.~\ref{eq:cprshortbalFB}. 
In the calculation, we could only match (approximately, as seen in panel (b)) the experimental values by multiplying by a factor $\eta$ the $\Delta^{\ast}$ expected from Eq.~17 in Ref.~\cite{Aminov1996}, with parameter $\bar{\tau}$, $\Delta_{\text{Al}}$ and $\gamma_B$ extracted from the experiments on Sample 3. As shown in Fig.~\ref{fig:activation_energy}(c) the parameter $\eta$ is relatively angle independent and close to 0.37, indicating that close to $T_c$ the induced gap is about one third of what expected by Eq.~17 of Ref.~\cite{Aminov1996}. On the one hand, this temperature regime is well above that explored in Refs.~\cite{baumgartner2020,baumgartner2021josephson}. On the other, the theory in Ref.~\cite{Aminov1996} is only valid far from $T_c$, therefore it does not apply to the measurements in Fig.~\ref{fig:activation_energy}(a). We have also tried to introduce a certain gaussian spread of the $E_J$ values in our model. However, even admitting a relatively large spread (standard deviation 25\% of the mean value), it is impossible to match the experimental data without a substantial reduction of $\Delta^{\ast}$ compared to the prediction of Eq.~17 in Ref.~\cite{Aminov1996}.



\section*{Conclusions}
In conclusion, we have studied the supercurrent diode effect and the magnetochiral anisotropy for the inductance in arrays of Josephson junctions with large spin-orbit coupling. These experiments complement those reported in the literature. We observe a dependence of the diode effect on the mutual orientation of supercurrent and lattice axes, which signals the presence of an additional Dresselhaus spin-orbit coupling term. 
Finally, we can correlate the anisotropy in the in-plane field dependence of the inductance with that of the phase-slip activation energy obtained from standard DC transport measurements. 
Superconducting diodes are the first step towards dissipation-free electronics. In perspective, they might play a crucial role in dissipationless memories, or in superconducting microwave detectors with ultra-high sensitivity.

\section*{Acknowledgments}
\begin{acknowledgments}
    We thank Hugues Pothier for helpful discussion.
    C.~B., L.~F., A.~C., S.~R., P.~E.~F.~J., D.~K., J.~F., N.~P., C.~S. acknowledge funding  by the Deutsche Forschungsgemeinschaft (DFG, German Research Foundation) – Project-ID 314695032 – SFB 1277 (Subprojects B05, B07, and B08). A.~C., P.~E.~F.~J., D.~K., J.~F. also benefited from the European Union’s Horizon 2020 research and innovation programme under Grant Agreement No.~881603~(Graphene Flagship Core~3). Work completed by S.~G., G.~C.~G., T.~L., M.~J.~M. is supported by Microsoft Quantum. 
\end{acknowledgments}
\vspace{2mm}

\bibliography{biblio}

\end{document}